\documentclass[letter,notitlepage,12pt]{revtex4}

\usepackage[]{graphicx}
\usepackage{amsmath}
\usepackage{cmap}

\newcommand{\Ethr}{\epsilon_{\text{thr.}}}
\newcommand{\meanLnA}{\left<\ln{A}\right>}
\newcommand{\tsectheta}{\times\sec{\theta}}
\newcommand{\rhos}{\rho_{s,600}}
\newcommand{\rhom}{\rho_{\mu,300}}
\newcommand{\Rm}{R_{\text{M}}}
\newcommand{\Xm}{x_{\text{max}}}

\begin{document}

\title{Muons in EAS with $E_0 \ge 10^{17}$~eV according to the data from Yakutsk array}

\author{A. V. Glushkov}
\author{A. Sabourov}
\affiliation{Yu. G. Shafer Institute of cosmophysical research and aeronomy}
\email{tema@ikfia.sbras.ru}

\begin{abstract}
  The lateral distribution of muons with threshold energy $\Ethr \simeq 1\tsectheta$~GeV have been studied in showers with energy $E_0 \ge 10^{17}$~eV. The data considered in the analysis have been collected from November 2011 to June 2013. Experimental values are compared to predictions obtained with the use of CORSIKA code within the framework of various hadron interaction models. The best agreement between experiment and theory is observed for QGSJETII-04 model. At $E_0 \simeq 10^{17}$~eV it complies to a mixed cosmic ray composition with the mean atomic number $\meanLnA \simeq 3.0 \pm 0.5$. At $E_0 \ge 4 \times 10^{17}$~eV the composition varies around the value $\meanLnA \simeq 0.5$.
\end{abstract}

\pacs{96.40.-z, 98.70.-f}

\keywords{ultra-high energy cosmic rays, mass composition}

\maketitle

\section{Introduction}

Ultra-high energy ($E \ge 10^{15}$~eV) cosmic rays (UHECR) are still remain a major scientific problem despite being studied worldwide by extensive air shower (EAS) arrays for good 50 years. Their mass composition is still not known exactly, and without this knowledge it is difficult to understand the character of nuclear interactions in this energy region. Muons with energy near $0.5 - 1.0$~GeV are very important component of EAS. They are poorly moderated in the atmosphere, are sensitive to the characteristics of nuclear interactions during development of a shower and to the chemical composition of cosmic rays (CR). Due to their yield and properties of lateral distribution they can be effectively registered with widely spaced ground arrays. Since 1978 Yakutsk EAS array has been continuously registering muons with the threshold energy $\Ethr \simeq 1.0\tsectheta$~GeV. During this period a large amount of experimental data has been accumulated. Analysis of this material~\citep{Astropart95, JETPl98, JETPl2000, YaF2000, YaF2005, JETP2006} has revealed that the development of showers with $E_0 \ge (3-5)\times 10^{18}$~eV differs significantly from those at lesser energies of CR. It also allowed us to estimate the fraction of primary gamma-quanta in the total CR flux at energies above $10^{17}$~eV~\citep{YakutskINR}.

\begin{figure}
  \centering
  \includegraphics[width=0.85\textwidth, clip]{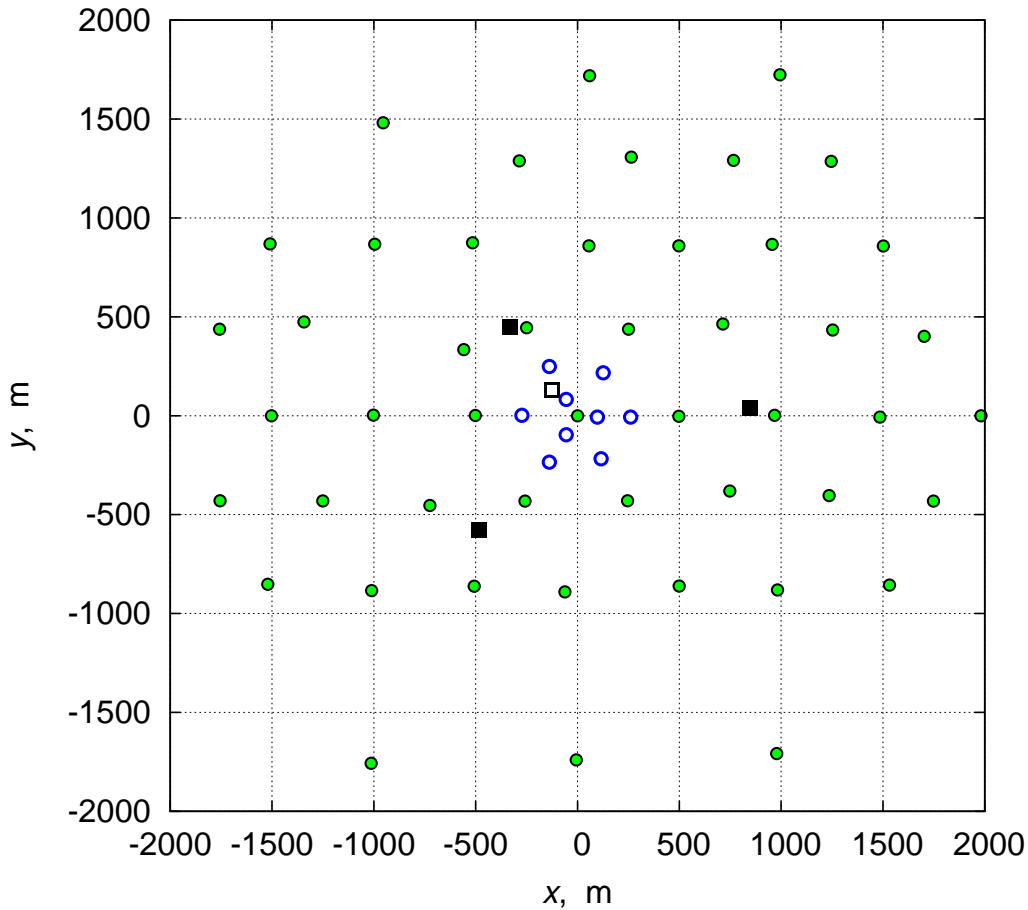}
  \caption{The layout of detectors location at Yakutsk array (since 1992). Green circles~--- detectors of the main array ($2\times 2$~m$^2$); empty blue circles~--- additional scintillation detectors ($2$~m$^2$); filled black squares~--- underground muon detectors of $10\times 2$~m$^2$ area and $1.0\times\sec{\theta}$~GeV threshold; empty black square~--- underground muon detector of $27\times 2$~m$^2$ and $0.5\times\sec{\theta}$~GeV threshold.}
  \label{fig:1}
\end{figure}

Recently we have shown \citep{JETPl2012} that the muon fraction in the total number of charged particles in EASs with energies $E_0 \ge 10^{17}$~eV changes significantly over periods of time. Until 1996, it fluctuated around a single stable position and then increased significantly. This was accompanied by near simultaneous variations in the energy spectrum and in the global anisotropy of CR within energy range $(1-10)\times 10^{17}$~eV~\citep{JETPl2012,AstroLet2013}. After 1996, during the next 7 years, the integral intensity of CR at $E_0 = 10^{17}$~eV increased by $(45 \pm 5)$\,\% and then started declining. As for the phase of the first harmonic $\phi_1 = 119^{\circ} \pm 18^{\circ}$ and its amplitude $A_1 = 0.030 \pm 0.014$ sampled during 1983-1994, they changed to values $\phi_1 = 284^{\circ} \pm 13^{\circ}$ and $A_1 = 0.033 \pm 0.010$ during 1998-2010. In recent years, a tendency has been manifested towards the change of these values in the opposite direction. It seems like the effect of some gargantuan explosion which have contributed a significant portion of heavy nuclei to the background. It's difficult to pinpoint exactly what kind of event in the Galaxy could led to such result. Here we need further studies involving temporal factor of the experimental data.

In this work we present the results of analyzes of muon data with $\Ethr \simeq 1.0\tsectheta$~GeV threshold collected during the period from November 2011 to June 2013. The geometry of the muon part of the array is shown on Fig.\ref{fig:1}. The work~\citep{JETPl2012} describes the technique of their control and calibration. The data from the detector with $\Ethr \simeq 0.5\tsectheta$~GeV~\citep{JETPl2013} are currently being accumulated and will be analyzed later.

\section{Results}

\begin{figure}
  \centering
  \includegraphics[width=0.85\textwidth]{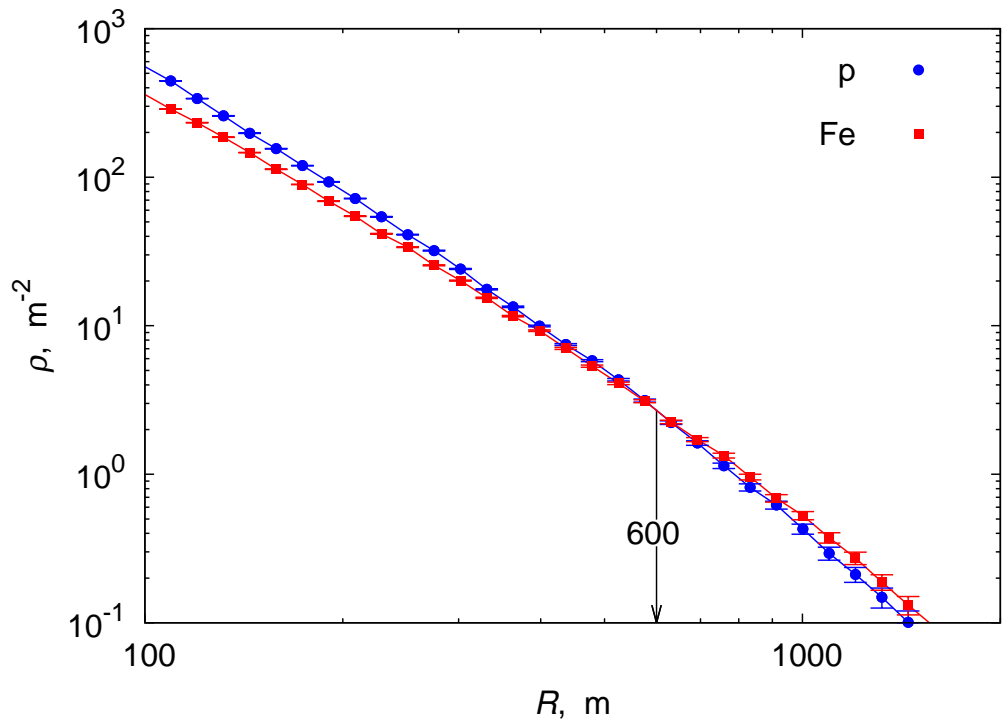}
  \caption{LDF of charged particles in showers with $E_0 = 10^{18}$~eV and $\cos{\theta} = 0.9$ from primary protons (blue circles) and iron nuclei (red squares) obtained within the framework of the QGSJETII-04~\cite{QGSJETII} model.}
  \label{fig:2}
\end{figure}

Further we consider EASs with zenith angles $\theta \le 45^{\circ}$ and axes fallen within a $1$~km radius circle in the center of the array and with the precision of axis detection no less than~$20$~m. The energy of primary particles was derived from relations:
\begin{equation}
  E_0\text{, eV} = (4.8 \pm 1.6) \times 10^{17} \cdot \rhos(0^{\circ})^{1.0 \pm 0.02}~\text{,}
  \label{eq:1}
\end{equation}
\begin{equation}
  \rhos(0^{\circ})\text{, m}^{-2} = \rhos(\theta) \cdot \exp{\frac{(\sec{\theta - 1}) \cdot 1020}{\lambda_{\text{p}}}} \text{}~\text{,}
  \label{eq:2}
\end{equation}
\begin{equation}
  \lambda_{\text{p}}\text{, g/cm}^2 = (450 \pm 44) + (32 \pm 15) \cdot \log_{10}{\rhos(0^{\circ})}~\text{,}
  \label{eq:3}
\end{equation}
where $\rhos(\theta)$ is the density of charged particles as measured by surface scintillation detectors at $R = 600$~m from a shower axis. The precision of $\rhos$ estimation in individual showers was no worse than 10\,\%. The relation~(\ref{eq:1}) unambiguously connects the $\rhos(0^{\circ})$ with $E_0$ at any given CR composition. It is due to the fact that at the distance $\sim 600$~m from the axis, the lateral distribution functions (LDF) of all charged particles inter-cross each other. It is demonstrated on Fig.\ref{fig:2} where two LDFs are shown, for charged particles in showers with $E_0 = 10^{18}$~eV and $\cos{\theta} = 0.9$ initiated by primary protons (blue circles) and iron nuclei (red squares) obtained with the use of QGSJETII-04 model~\citep{QGSJETII}. Values for $\rhos(\theta)$ were derived from the modified Linsley approximation~\citep{Linsley62}:
\begin{equation}
  f_{s}(R,\theta) = \rhos(\theta) \cdot \frac{600}{R} \cdot
  \left(
    \frac{\Rm + 600}{\Rm + R}
  \right)^{b_s - 1}\text{,}
  \label{eq:4}
\end{equation}
where $\Rm$ is the Molier radius which depends on air temperature ($T, ^\circ$\,C) and pressure ($P$, mbarn):
\begin{equation}
  \Rm\text{, m} \simeq \frac{7.5 \times 10^4}{P} \cdot \frac{T}{273}~\text{.}
  \label{eq:5}
\end{equation}
The value for $\Rm$ is measured in each individual event (for Yakutsk $\left<T\right> \simeq -18^{\circ}$\,C, $\left<\Rm\right> \simeq 70$~m). In the expression (\ref{eq:4}) $b_s$ is the parameter defined in~\citep{Yakutsk76}:
\begin{equation}
  b_s = 1.38 + 2.16 \times \cos{\theta} + 0.15 \times \log_{10}{\rhos(\theta)}\text{.}
  \label{eq:6}
\end{equation}

\begin{figure}
  \centering
  \includegraphics[width=0.65\textwidth]{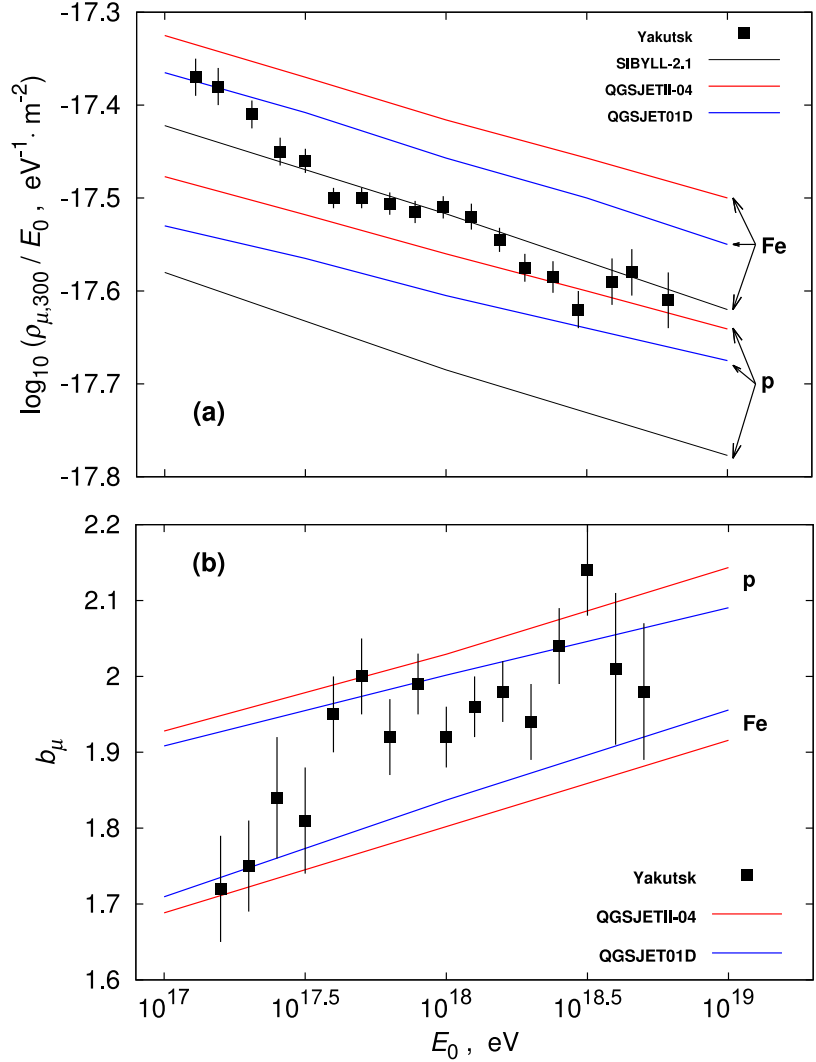}
  \caption{{\bf (a)} Muon densities at 300~m from shower axis normalized to $E_0$; {\bf (b)} the steepness parameter $b_{\mu}$ of muon LDF ($\Ethr \simeq 1.0 \tsectheta$~GeV). On both pictures theoretical predictions for primary protons (p) and iron nuclei (Fe) are represented with red (QGSJETII-04), blue (QGSJET01D) and black (SIBYLL-2.1) lines. Squares~--- experimental values.}
  \label{fig:3}
\end{figure}

On Fig.\ref{fig:3}{\bf (a)} with black squares are shown muon densities at the distance $300$~m from axis of showers within a considered group, with mean values of $E_0$ and $\cos{\theta} = 0.9$. Normalization to primary energy $\log_{10}\left({\left<\rhom\right> / \left<E_0\right>}\right)$ gives a representation of muon data that is more descriptive and convenient for further analysis. Mean LDFs were obtained within energy bins with logarithmic step $h = \Delta\log_{10}{E_{0}} = 0.2$ which were subsequently shifted towards higher energies by $0.5\,h$. This procedure was performed for a detailed test of an agreement between the experiment and various hadron interaction models. $\rhom$ values were obtained from approximations of mean LDFs. When constructing an LDF, muon densities were multiplied by normalizing ratio $\left<E_0\right> / E_0$ and averaged over an energy cut in radial bins $\Delta\log_{10}{R} = 0.04$. Mean muon densities were determined from the expression
\begin{equation}
  \left<\rho_{\mu}(R_i)\right> =
  \frac{\sum_{n = 1}^{N_1} \rho_{\mu}(R_i)}{N_1 + N_0}\text{,}
  \label{eq:7}
\end{equation}
where $N_1$ and $N_0$ are the numbers of operated muon detectors at axis distances within the interval $(\log_{10}(R_i), \log_{10}(R_i) + \Delta\log_{10}{R})$. The indexes denote whether a detectors had non-zero $(N_1)$ or zero $(N_0)$ readings during the registration of event. Zero readings are related to cases when a detector hasn't registered any muons while being in a wait state. Mean LDFs were approximated according to functions~\citep{YaF2000}:
\begin{equation}
  \rho_{\mu} = f_{\mu} \cdot \left(1 + \frac{R}{2000}\right)^{-6.5}
  \label{eq:8}
\end{equation}
with well-known relation by \citet{Greisen60}:
\begin{equation}
  f_{\mu}(R,\theta) = \rho_{\mu,600}(\theta) \cdot \left(\frac{600}{R}\right)^{0.75}
  \cdot \left(\frac{R_0 + 600}{R_0 + r}\right)^{b_{\mu} - 0.75}\text{,}
  \label{eq:9}
\end{equation}
where $R_0=280$~m, $b_{\mu}$ is a free parameter. The best fit values of $b_s$, $\rhos(\theta)$ in (\ref{eq:4}) and $b_{\mu}$, $\rho_{\mu,600}(\theta)$ in (\ref{eq:9}) were determined with the use of $\chi^2$ minimization. Error bars on Fig.\ref{fig:3}{\bf (a)} include the entire combination originated from statistics of events and averaging of the data. Lines represent expected values predicted by hadron interaction models QGSJETII-04 (red), QGSJET01D~\cite{QGSJET} (blue) and SIBYLL-2.1~\cite{SIBYLL} (black). Simulations were performed with the use of CORSIKA code~\cite{CORSIKA} (version 6.990 in the case of SIBYLL-2.1 and QGSJET01D and 7.3700 in the case of QGSJETII-04). 200 showers were simulated per each set of initial shower parameters (mass of primary particle, energy and zenith angle). To speed-up the computations, the thin-sampling algorithm was activated in the CORSIKA code with the parameters $E_i / E_0 \in [3.16 \times 10^{-6}, 10^{-5}]$ and $w_{\text{max}} \in [10^4, 3.16 \times 10^{6}]$ depending on the primary energy~\cite{CORSIKAUG}. The density was calculated directly from total number of particles arrived at a detector of given area.

It is clearly seen that the experiment is not consistent with SIBYLL at neither given composition; the model predicts significantly less muon yield. Other two models agree with our experiment much better and allow to estimate the mass composition of primary particles. To simplify, let us consider a two-component composition, consisting of protons and iron nuclei. In this case the relation
\begin{equation}
  \meanLnA = W_{\text{p}} \cdot \ln{1} + W_{\text{Fe}} \cdot \ln{56}
  \label{eq:10}
\end{equation}
gives weighting functions $W_{\text{p}} = 1 - W_{\text{Fe}}$ and $W_{\text{Fe}} = \meanLnA / \ln{56}$. Within the framework of this hypothesis, according to the QGSJET01D model we have:
\begin{equation}
  W_{\text{Fe}} = \frac{d_{\text{exp}} - d _{\text{p}}}
  {d_{\text{Fe}} - d_{\text{p}} }~\text{,}
  \label{eq:11}
\end{equation}
where $d = \log_{10}\left({\rhom} / E_0\right)$~--- are the values obtained in the experiment (exp) and in simulation.

\begin{figure}
  \centering
  \includegraphics[width=0.85\textwidth]{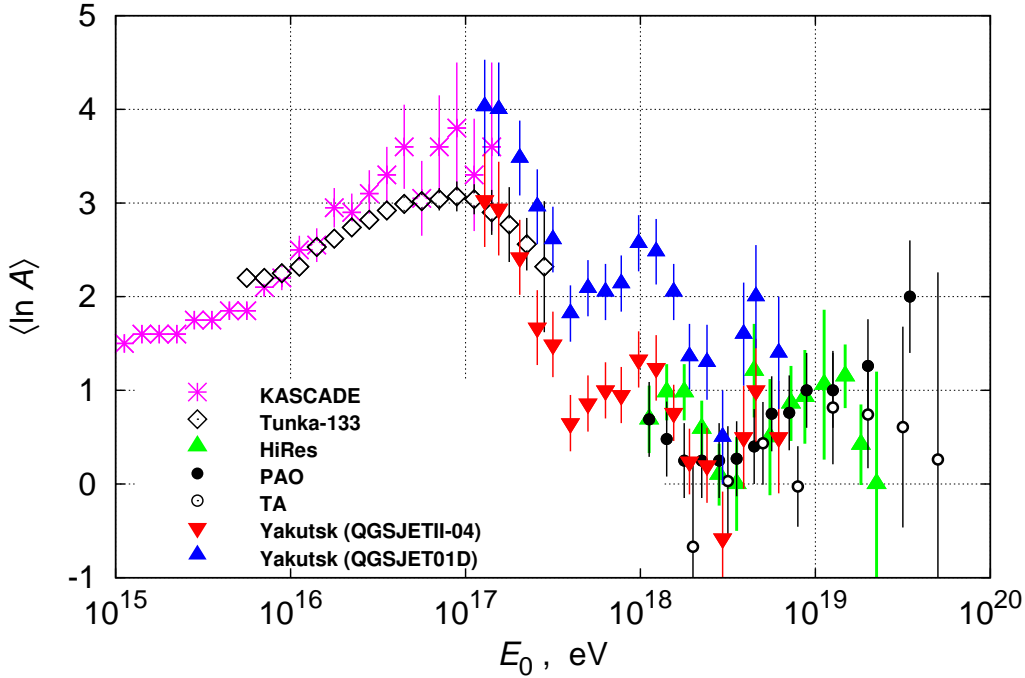}
  \caption{Mean atomic number of CR versus the energy of primary particles according to various experiments. Purple stars~--- KASCADE, white diamonds~--- Tunka-133, green triangles~--- HiRes, black circles~--- PAO, white circles~--- TA. Red triangles~--- estimations derived from the data of Yakutsk experiment within the framework of QGSJETII-04; blue triangles~--- estimations obtained from the same data within the framework of QGSJET01D.}
  \label{fig:4}
\end{figure}

With red and blue triangles on Fig.\ref{fig:4} are shown energy dependencies of CR mass composition according to predictions of QGSJETII-04 and QGSJET01D correspondingly. For comparison, on the same figure the data from various EAS experiments are shown. Purple stars denote the results of KASCADE obtained during the period from May 1998 to December 1999~\cite{Ulrich2001}. White diamonds represent the data of Tunka-133 obtained from the LDF of Cherenkov radiation emitted by EAS during two winter observational periods (2009 - 2011)~\cite{Tunka2011}. Other values were derived from $\Xm(E_0)$ dependence established in experiments and simulations within the framework of QGSJET01 model, with the use of the expression (\ref{eq:11}) with substitution $d = \Xm$. With green triangles are shown HiRes data related to the observational period from November 1999 to September 2001~\cite{HiRes2005}. Black circles~--- results from PAO obtained between December 2004 and September 2010~\cite{PAO2011}, open circles~--- interpretation of the Telescope Array data within the framework of QGSJETII model.~\cite{TA2013}.

\section{Discussion}

\begin{figure}
  \centering
  \includegraphics[width=0.85\textwidth]{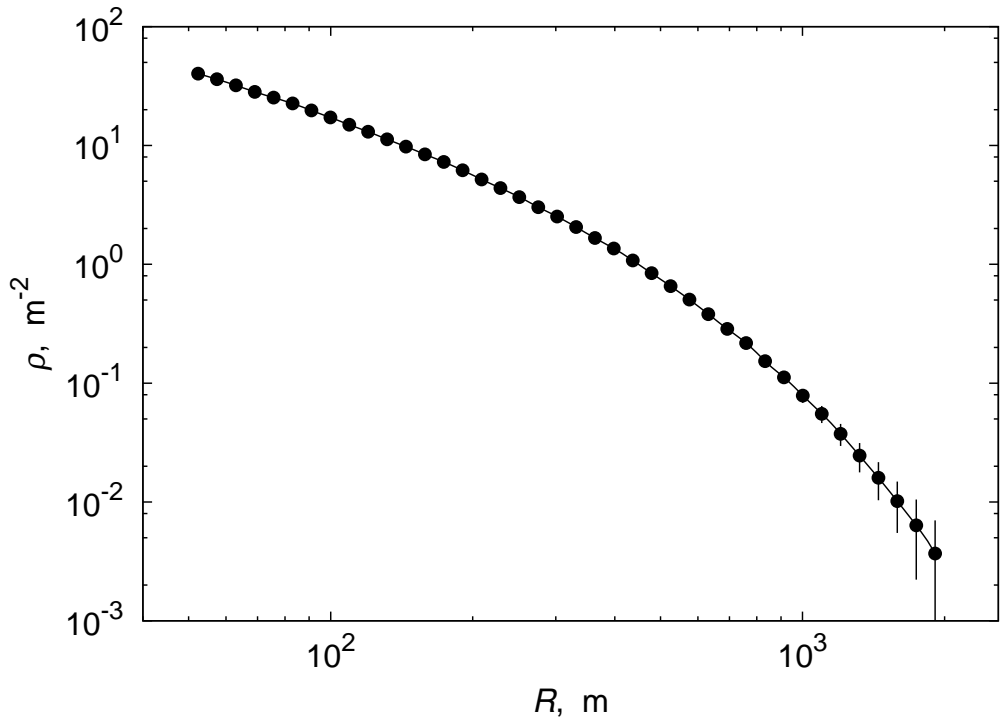}
  \caption{LDF of muons with $1.0\tsectheta$~GeV threshold obtained with the use of QGSJET01D model for primary protons with primary energy $10^{18}$~eV at $\cos{\theta} = 0.9$. The line represents approximation (\ref{eq:9}) with $b_{\mu} = 2.0$.}
  \label{fig:5}
\end{figure}

Results presented on Fig.\ref{fig:4} give evidence that in energy region $(1 - 5) \times 10^{17}$~eV the CR composition probably changes rapidly towards lighter nuclei. Our data do not contradict this scenario. It is seen that QGSJETII-04 model better agrees with the experiment than QGSJET01D, especially in view of the works~\cite{JETPl2012, AstroLet2013}, which have demonstrated that at $E_0 \le 10^{18}$~eV the mass composition changes rapidly. Yet, it is still a bit premature to make a strong conclusion~--- both models predict similar shapes of muon LDFs and agree with experiment, as evidenced by the parameter of LDF's steepness $b_{\mu}$ shown on Fig.\ref{fig:3}{\bf (b)}. Experimental values of this parameter confirm the above-mentioned hypothesis about lightening of the CR mass composition with the increase of primary energy. The LDFs obtained in simulations are well-described by the function (\ref{eq:8}) in a wide range of distances from a shower axis. It is demonstrated on Fig.\ref{fig:5} where points represent LDF for muons with $1.0\tsectheta$~GeV threshold obtained with the use of QGSJET01D model for primary protons with $E_0 = 10^{18}$~eV and $\cos{\theta} = 0.9$. Line is the approximation of (\ref{eq:9}) with steepness parameter $b_{\mu} = 2.00 \pm 0.01$ found with weighted least square method. Similar picture is observed at other values of initial shower parameters (i.e. energy, zenith angle and composition).

\begin{figure}
  \centering
  \includegraphics[width=0.85\textwidth]{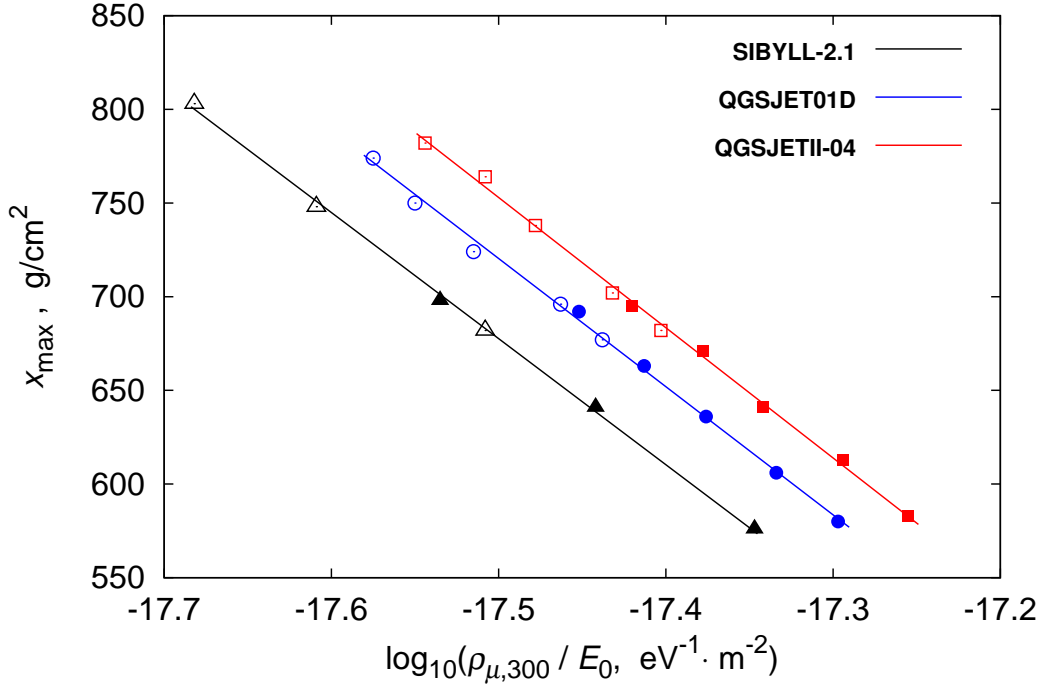}
  \caption{The dependence between $\Xm$ and muon density (at $\Ethr = 1.0\tsectheta~\text{GeV}$) normalized to primary energy ($\rhom / E_0$). Predictions of QGSJETII-04 (red squares), QGSJET01D (blue circles) and SIBYLL-2.1 (black triangles) for primary protons (open symbols) and iron nuclei (filled symbols) within the energy range $10^{17} - 10^{19}$~eV at $\cos{\theta} = 0.9$. Lines denote linear approximations.}
  \label{fig:6}
\end{figure}

\begin{figure}
  \centering
  \includegraphics[width=0.85\textwidth]{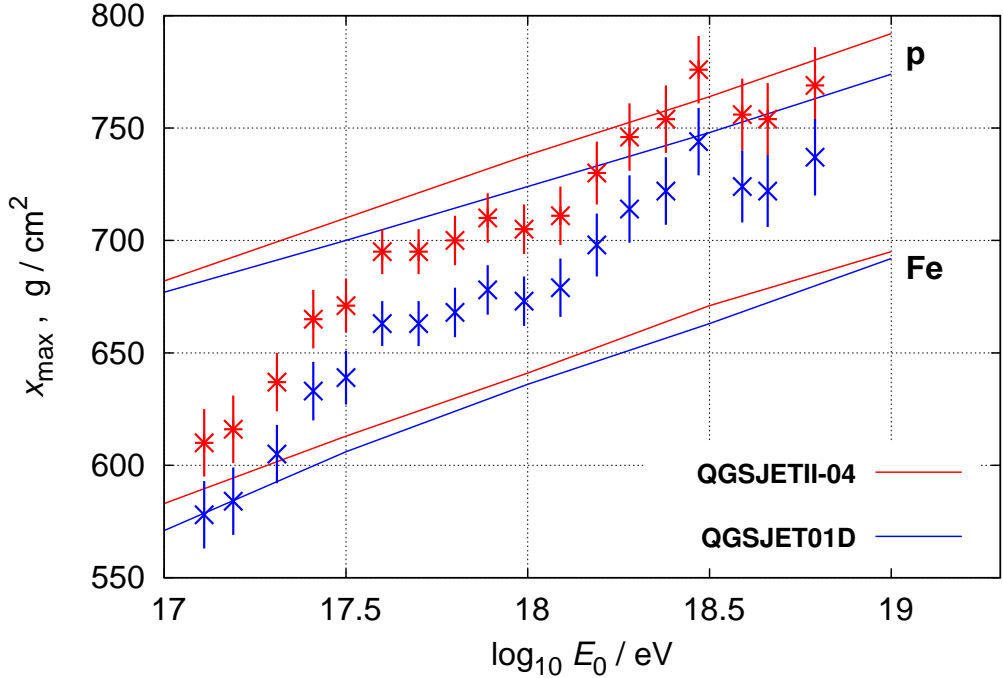}
  \caption{Energy dependence of $\Xm$ according to QGSJETII-04 (red lines) and QGSJET01D (blue line) for primary protons (p) and iron nuclei (Fe). Red stars and blue crosses represent the interpretation of experimental data within the frameworks of the two models correspondingly.}
  \label{fig:7}
\end{figure}

Our estimations of $\meanLnA$ based on the (\ref{eq:10}) do not differ drastically from other, more traditional method based on the $\Xm$ (see e.g.,~\cite{KampertUnger2011}):
\begin{equation}
  \meanLnA = \frac{\left<\Xm^{\text{p}}\right> - \left<\Xm^{\text{exp.}}\right>}
                  {\left<\Xm^{\text{p}}\right> - \left<\Xm^{\text{Fe}}\right>}
                  \cdot \ln{56}\text{.}
  \label{eq:12}
\end{equation}
It is due to some peculiarity of EAS muons, which is demonstrated on Fig.\ref{fig:6}. It shows that within the framework of any shower development model, between the depth of maximum $\Xm$ and the logarithm of muon density normalized to $E_0$, there is a quasi-linear dependence at any composition of primary particles:
\begin{equation}
  \Xm = A + C \cdot \ln{\frac{\rho_{\mu}}{E_0}}\text{.}
  \label{eq:13}
\end{equation}
If one inserts (\ref{eq:13}) into (\ref{eq:12}), then (\ref{eq:10}) follows. The relations visualized on Fig.\ref{fig:6} allow one to find $\left<\Xm\right>$ from muon density normalized to primary energy. In our case, from the plot on this figure follow the dependencies shown on Fig.\ref{fig:7}. This method of extracting the depth of maximum from muon data, in our opinion, is comfortably simple and efficient from practical point of view. With sufficient statistics it is essentially a fully functional alternative to a technique of $\left<\Xm\right>$ determination from the LDF of Cherenkov light radiation. It is possible to use different muon thresholds and distances from shower axis, where muon density is sampled.

\section{Conclusion}

The comparison of the Yakutsk array data to modern ultra-high energy interaction models has demonstrated once again the importance that this component presents for studying of shower development and CR mass composition. The results from Fig.\ref{fig:3} demonstrate a certain degree of agreement between the experiment and QGSJETII-04 and QGSJET01D models on the whole energy range $E_0 \simeq 10^{17} - 10^{19}$~eV. Estimations of the mean CR mass composition from Fig.\ref{fig:4}  obtained within the frameworks of these models are close enough to worldwide data~\cite{Ulrich2001, Tunka2011, HiRes2005, PAO2011}. The QGSJETII-04 model agrees with them better than QGSJET01D. Our data displayed on Fig.\ref{fig:4} have added to the whole picture that testifies of a rapid change in the CR composition towards lighter nuclei in energy range $E_0 \simeq (1 -5) \times 10^{17}$~eV. It cannot be ruled out that at $E_0 \simeq 10^{18}$ there is a peak of local heaviness of CR composition. It is seen from Fig.\ref{fig:5} that LDFs obtained from simulations are well-described by the approximation (\ref{eq:8}) within a wide range of distances from shower axis. This approximation is convenient for comparison between the experimental data and theoretical predictions as seen on Fig.\ref{fig:3}{\bf (b)}. Fig.\ref{fig:6} suggests that muon component contains some possibilities yet to discover. In particular, it allows to relatively easy determine the depth of maximum EAS development (see Fig.\ref{fig:7}). We suppose that muons can play a vital role in the energy cross-calibration between world's EAS arrays, where there is still no clarity in relation to correctness of any given method of estimating the energy of primary particle.

\acknowledgements
The work was conducted with the financial support from Russian Academy of Science within the framework of research program ``Fundamental properties of matter and Astrophysics'' and is supported by RFBR grant \#\,13--02--12036 ofi-m-2013.

\end{document}